\documentclass{article}
\def\Ref#1{(\ref{#1})}
\usepackage{amsmath}
\usepackage{amssymb}
\usepackage{cite}

\def\d{{\rm d}}
\begin{document}
\begin{titlepage}

\begin{center}
{\Large\textbf{Phase transition in annihilation-limited
processes}}

\vskip 2 cm

{Mohammad~Khorrami{\footnote {mamwad@mailaps.org}} \&
Amir~Aghamohammadi{\footnote {mohamadi@alzahra.ac.ir}}} \vskip 5
mm

\textit{  Department of Physics, Alzahra University,
             Tehran 1993891167, Iran. }
\end{center}

\begin{abstract}
\noindent A system of particles is studied in which the stochastic
processes are one-particle type-change (or one-particle diffusion)
and multi-particle annihilation. It is shown that, if the
annihilation rate tends to zero but the initial values of the
average number of the particles tends to infinity, so that the
annihilation rate times a certain power of the initial values of
the average number of the particles remain constant (the double
scaling) then if the initial state of the system is a
multi-Poisson distribution, the system always remains in a state
of multi-Poisson distribution, but with evolving parameters. The
large time behavior of the system is also investigated. The system
exhibits a dynamical phase transition. It is seen that for a
$k$-particle annihilation, if $k$ is larger than a critical value
$k_{\mathrm{c}}$, which is determined by the type-change rates,
then annihilation does not enter the relaxation exponent of the
system; while for $k < k_{\mathrm{c}}$, it is the annihilation (in
fact $k$ itself) which determines the relaxation exponent.

\end{abstract}
\begin{center} {\textbf{PACS numbers:}} 05.40.-a, 02.50.Ga

{\textbf{Keywords:}} reaction-diffusion, annihilation,
diffusion-limited, phase transition
\end{center}
\end{titlepage}
\section{Introduction}
People have studied reaction-diffusion systems, using analytical
techniques, approximation methods, and simulations. A large
fraction of exact results belong to low-dimensional (specially
one-dimensional) systems, as solving low-dimensional systems
should in principle be easier
\cite{ScR,ADHR,KPWH,HS1,PCG,HOS1,HOS2,AL,AKK,RK2,AKK2,AM1,KA}.
Despite their simplicity, these systems exhibit rather rich and
non-trivial dynamical and stationary behavior. Studies on the
models  far from equilibrium have shown that there is a remarkably
rich variety of critical phenomena\cite{ScR}. Among the important
aspects of reaction-diffusion systems, are the stationary state of
the system (or one of the quantities describing the system) and
the relaxation behavior of the system towards this configuration.

Field theoretic methods have been used to study diffusion,
recombination, and other dynamic manifestations of
non-quantum-mechanical objects (for a review see \cite{MG}). The
techniques of field theory on a lattice are used to examine the
diffusion and reaction processes of particles \cite{MG}. The field
theoretic methods and the dynamic renormalization group (RG) have
been applied to study the universal scaling properties of
reaction-diffusion models \cite{THV,Ta}. They have also been used
to study fluctuations in reaction-diffusion problems, for example
to study the single-species annihilation of $k$ particles to $l$
particles ($l < k$) \cite{THV}. In \cite{BH} a model is
investigated where organisms die and give birth with equal rates,
and also diffuse. It is shown there that at dimensions smaller
than or equal to 2, there is aggregation, while at larger
dimensions there is no clustering. In \cite{MPGS,BHPR}, a model is
investigated where particles are created, annihilated, and
diffused on a lattice. For the case of one- or two-particle
annihilation and creation, exact results are obtained. In all
these cases, bosonic formulations have been used, meaning that
each site can be occupied by more than one particle.

In this paper, a stochastic model is considered in which the
(stochastic) variables of the system are the numbers of various
types of particles. The term {\em type} can refer to species as
well as position of the particle. There are single-particle
type-changes, as well as $k$-particle annihilations. Specially, a
case is studied where the annihilation rate tends to zero while
this rate times the $(k-1)$'th power of the initial values of the
average particle-numbers remain constant. It is shown that in this
double-scaling limit, the evolution equations for the annihilation
operators contain only annihilation operators. Specifically, if
the initial state of the system is a multi-Poisson distribution,
then the system always remains in a state of multi-Poisson
distribution, but with evolving parameters. The parameters evolve
in a set of equations which in fact are the mean field equations.

It is further shown that the system exhibits a dynamical phase
transition. The large time behavior of the system is controlled by
the spectrum of the evolution operator corresponding to
single-particle type-changes, and $k$. It is shown that if $k$
exceeds a critical value $k_{\mathrm{c}}$, which is determined by
type-change rates, annihilation does not enter the relaxation
exponent of the system; while for $k < k_{\mathrm{c}}$, it is the
annihilation which determines the relaxation exponent.

The scheme of the paper is the following. In section 2, some
general techniques are introduced, mainly to fix notation. In
section 3, the double-scaling is discussed. In section 4, the
large-time behavior of the system and the dynamical phase
transition are investigated. Section 5 is devoted to the
concluding remarks.
\section{The general technique}
To fix notation, let's briefly introduce the general technique.
Consider a system the state of which is characterized by a set of
nonnegative integers (say the number of different particles). The
vector space corresponding to such a system can be constructed
using the raising and lowering operators ($A_\mu^\dagger$'s and
$A^\mu$'s, respectively), and the corresponding number operators
$N^\mu$, through
\begin{align}\label{a1}
[A^\mu,A^\dagger_\nu]=&\delta^\mu_\nu,\nonumber\\
[A^\mu,A^\nu]=&0,\nonumber\\
[A^\dagger_\mu,A^\dagger_\nu]=&0,\nonumber\\
A^\mu\,|0\rangle=&0,\nonumber\\
N^\mu:=&
\Delta^{\mu\,\alpha}{}_\beta\,A^\dagger_\alpha\,A^\beta,\nonumber\\
\Delta^{\mu\,\alpha}{}_\beta:=&\delta^\mu_\beta\,\delta^\alpha_\beta,\nonumber\\
|n\rangle:=& \prod_\mu(A_\mu^\dagger)^{n_\mu}\,|0\rangle,\nonumber\\
\langle m|n\rangle=&\delta_{m\,n},
\end{align}
where the kets $|n\rangle$ form a basis for the vector space
corresponding to the system. Note that by several types of
particles, one may mean several species of particles or particles
in several places (or both).

Corresponding to any set of probabilities $P_n$ of finding the
system in the state $|n\rangle$ (having $n_\mu$ particles of type
$\mu$), there is a probability vector in the vector space which is
\begin{equation}\label{a2}
|P\rangle=\sum_n P_n\,|n\rangle.
\end{equation}

Any physical state of such a system is characterized by a vector
like
\begin{equation}\label{a3}
|P\rangle=f(A^\dagger)\,|0\rangle,
\end{equation}
where $f(A^\dagger)$ is a Taylor series in $A^\dagger$ with
nonnegative coefficients and with the sum of coefficients equal to
one. This last condition can be written as one of the following
equivalent forms
\begin{align}\label{a4}
f(S)=&1,\nonumber\\
\langle\mathcal{S}|\,f(A^\dagger)=&\langle\mathcal{S}|,
\end{align}
where $S$ is a covector all of its coefficients are equal to one,
and
\begin{equation}\label{a5}
\langle\mathcal{S}|=\langle 0|\,e^{S_\mu\,A^\mu}.
\end{equation}

The observables of such a system are functions of the number
operators. The expectation value of the observable $g(N)$ is
\begin{align}\label{a6}
\langle g(N)\rangle&=\langle\mathcal{S}|\,g(N)\,|P\rangle,
\nonumber\\
&=\langle 0|\,g(N+A)\,f(A^\dagger+S)\,|0\rangle,\nonumber\\
&=:\langle 0|\,\bar g(A)\,f(A^\dagger+S)\,|0\rangle.
\end{align}
where in the last equality commutation relations between $A$'s and
$A^\dagger$'s have been used to rearrange them in $g(N+A)$ so that
$A^\dagger$'s are all in the left of $A$'s. Specially, if the
system has a multi-Poisson probability distribution with
parameters $\Lambda^\mu$:
\begin{equation}\label{a7}
|P\rangle=e^{\Lambda^\mu\,(A^\dagger_\mu-S_\mu)}\,|0\rangle,
\end{equation}
then
\begin{equation}\label{a8}
\langle g(N)\rangle=\bar g(\Lambda).
\end{equation}

A general continuous-time stochastic process is described by a
linear operator (Hamiltonian) $H$ with nonnegative nondiagonal
elements and the property that
\begin{equation}\label{a9}
\langle\mathcal{S}|\,H=0.
\end{equation}
Such a Hamiltonian can be written in terms of the annihilation and
creation operators. Specifically, a process involving the
annihilation of $k$ particles and creation of $l$ particles is
described by the Hamiltonian
\begin{align}\label{a10}
H=&(A^\dagger_{\alpha_1}\cdots A^\dagger_{\alpha_l}\,
A^{\beta_1}\cdots A^{\beta_k}\nonumber\\
&- S_{\alpha_1}\cdots S_{\alpha_l}\,
\Delta^{\beta_1\,\gamma_1}{}_{\delta_1}\cdots
\Delta^{\beta_k\,\gamma_k}{}_{\delta_k}\,
A^\dagger_{\gamma_1}\cdots A^\dagger_{\gamma_k}\,
A^{\delta_1}\cdots A^{\delta_k})\,
C^{\alpha_1\cdots\alpha_l}{}_{\beta_1\cdots\beta_k}(N),\nonumber\\
=&[A^\dagger_{\alpha_1}\cdots A^\dagger_{\alpha_l}\,
A^{\beta_1}\cdots A^{\beta_k} -S_{\alpha_1}\cdots S_{\alpha_l}\,
\mathcal{N}(N^{\beta_1}\cdots N^{\beta_k})]\,
C^{\alpha_1\cdots\alpha_l}{}_{\beta_1\cdots\beta_k}(N),
\end{align}
where $C$'s are nonnegative rates, and $\mathcal{N}$ means
normal-ordering, that is putting $A^\dagger$'s at the left of
$A$'s.

The evolution of the state vector of the system ($|P(t)\rangle$)
is through
\begin{equation}\label{a11}
|P(t)\rangle= U(t,0)\,|P(0)\rangle,
\end{equation}
where
\begin{align}\label{a12}
\frac{\partial}{\partial t}U(t,0)=&
H\,U(t,0),\nonumber\\
U(0,0)=& 1.
\end{align}
So the expectation value of an observable $Q$ at the time $t$ can
be written like
\begin{align}\label{a13}
\langle Q(t)\rangle=&\langle\mathcal{S}|\,Q\,|P(t)\rangle,\nonumber\\
=&\langle\mathcal{S}|\,Q^{\mathrm{H}}(t)\,|P(0)\rangle,
\end{align}
where
\begin{align}\label{a14}
Q^{\mathrm{H}}(t):=&U^{-1}(t,0)\,Q\,U(t,0),\nonumber\\
\frac{\d}{\d t} Q^{\mathrm{H}}(t)=&[Q^{\mathrm{H}}(t),
H^{\mathrm{H}}(t)],
\end{align}
One notes that the Heisenberg operators $Q^{\mathrm{H}}$ are in
fact the ordinary operators in them $A$'s and $A^\dagger$'s are
substituted by $A(t)$'s and $A^\dagger(t)$'s.

One also has
\begin{align}\label{a15}
\langle Q(t)\rangle=&\langle 0|\,\tilde Q\,|\tilde P(t)\rangle,\nonumber\\
=&\langle 0|\,\tilde Q^{\mathrm{H}}(t)\,|\tilde P(0\rangle,
\end{align}
where
\begin{align}\label{a16}
|\tilde P\rangle:=&e^{S_\mu\,A^\mu}\,|P\rangle,\nonumber\\
\tilde Q:=&e^{S_\mu\,A^\mu}\,Q\,e^{-S_\mu\,A^\mu},\nonumber\\
\tilde Q^{\mathrm{H}}(t):=&\tilde U^{-1}(t,0)\,\tilde Q \,\tilde
U(t,0),
\end{align}
and $\tilde U$ is defined similar to \Ref{a12}, but with $\tilde
H$ in place of $H$. It is seen that the effect of \textit{tilde}
on the operators is just to change $A^\dagger_\alpha$ to
$(A^\dagger_\alpha+S_\alpha)$.
\section{Double scaling in annihilation processes of low rates}
Consider a reaction-annihilation process with the Hamiltonian
\begin{equation}\label{a32}
H=H_0+H_{\mathrm{I}},
\end{equation}
where
\begin{align}\label{a33}
H_0:=&M^{\alpha}{}_\beta\,A^\dagger_\alpha\,A^\beta,\\
\label{a34} H_{\mathrm{I}}:=&\sum_k
C_{\beta_1\cdots\beta_k}\,[A^{\beta_1}\cdots
A^{\beta_k}-\mathcal{N}(N^{\beta_1}\cdots N^{\beta_k})],
\end{align}
where
\begin{equation}\label{a35}
S_\alpha\,M^\alpha{}_\beta=0.
\end{equation}
Since only the symmetric part of $C$ enters the Hamiltonian, from
now on it is assumed that $C$ is symmetric. $H_0$ describes a
reaction, change of a particle of type $\beta$ to a particle of
type $\alpha$ with the rate $M^\alpha{}_\beta$, while
$H_{\mathrm{I}}$ describes annihilations. For the observable
$g(N)$, one has then
\begin{equation}\label{a36}
\langle g[N(t)]\rangle=\langle 0| \tilde U^{-1}(t,0)\, \bar
g(A)\,\tilde U(t,0)\,|\tilde P(0)\rangle,
\end{equation}
where the evolution of $\tilde U$ is governed by by $\tilde H$:
\begin{equation}\label{a37}
\tilde H=M^{\alpha}{}_\beta\,A^\dagger_\alpha\,A^\beta +\sum_k
C_{\beta_1\cdots\beta_k}\,\{A^{\beta_1}\cdots
A^{\beta_k}-\mathcal{N}[(N^{\beta_1}+A^{\beta_1})\cdots
(N^{\beta_k}+A^{\beta_k})]\},
\end{equation}
and
\begin{equation}\label{a38}
|\tilde P(0)\rangle =\tilde f(A^\dagger)\,|0\rangle.
\end{equation}
Suppose that the initial state vector describes a large number of
particles, and the annihilation rates are small, specifically so
that $\tilde f(A^\dagger/\lambda)$ and
$(\lambda^{k-1}\,C_{\beta_1\cdots\beta_k})$ (for all $k$'s) both
exist as $\lambda\to\infty$. One can then define another pair of
annihilation and creation operators through the transformation
\begin{align}\label{a39}
A^\beta=:\lambda\,a^\beta,\nonumber\\
A^\dagger_\alpha=:\lambda^{-1}\,a^\dagger_\alpha.
\end{align}
Writing $A$'s and $A^\dagger$'s in terms of $a$'s and
$a^\dagger$'s, and sending $\lambda$ to infinity, it is seen that
in the Hamiltonian $\tilde H$ only those terms survive that are
linear in $A^\dagger$. So in this double-scaling limit, one can
use instead of $\tilde H$ the Hamiltonian $\tilde H_{\mathrm{s}}$:
\begin{equation}\label{a40}
\tilde H_{\mathrm{s}}:=M^{\alpha}{}_\beta\,A^\dagger_\alpha
\,A^\beta -\sum_k
k\,C_{\beta_1\cdots\beta_k}\,N^{\beta_1}\,A^{\beta_2}\cdots
A^{\beta_k},
\end{equation}
where use has been made of the fact that $C$'s are symmetric. It
is now easily seen that in this limit,
\begin{equation}\label{a41}
\langle g[N(t)]\rangle=\langle 0|\, \bar g[\tilde
A_{\mathrm{s}}(t)]\,|\tilde P(0)\rangle,
\end{equation}
where $\tilde A^\alpha_{\mathrm{s}}(t)$'s satisfy
\begin{equation}\label{a42}
\frac{\d}{\d t}\,\tilde
A^\alpha_{\mathrm{s}}=\left(M^\alpha{}_\beta
-\Delta^{\beta_1\,\alpha}{}_\beta\,\sum_k
k\,C_{\beta_1\cdots\beta_k} \,\tilde
A^{\beta_2}_{\mathrm{s}}\cdots \tilde
A^{\beta_k}_{\mathrm{s}}\right) \,\tilde A^\beta_{\mathrm{s}}.
\end{equation}
This is a set of differential equations for $\tilde
A^\alpha_{\mathrm{s}}$'s, which are commuting at $t=0$ and remain
commuting at later times. Specifically, if the initial state of
the system is a multi-Poisson distribution \Ref{a7}, then with
this evolution the system always remains in a state of
multi-Poisson distribution, but with evolving parameters
$\Lambda^\alpha(t)$ which satisfy
\begin{equation}\label{a43}
\frac{\d}{\d t}\,\Lambda^\alpha=\left(M^\alpha{}_\beta
-\Delta^{\beta_1\,\alpha}{}_\beta\,\sum_k
k\,C_{\beta_1\cdots\beta_k}
\,\Lambda^{\beta_2}\cdots\Lambda^{\beta_k}\right)\,\Lambda^\beta.
\end{equation}

This equation can be solved perturbatively. One defines
\begin{equation}\label{a44}
\Lambda^\alpha(t)=:R^\alpha{}_\beta(t)\,Y^\beta(t),
\end{equation}
where
\begin{align}\label{a45}
\frac{\d}{\d t}\,R=&M,\nonumber\\
R(0)=&1.
\end{align}
To be more specific, let's consider a system described by a
reaction and a $k$-particle annihilation. One has
\begin{equation}\label{a46}
\frac{\d}{\d t}\,Y^\alpha=(R^{-1})^\alpha{}_\sigma\,
\,D^\sigma{}_{\beta_1\cdots\beta_k}\,(R^{\beta_1}{}_{\alpha_1}\,Y^{\alpha_1})
\cdots(R^{\beta_k}{}_{\alpha_k}\,Y^{\alpha_k}),
\end{equation}
where
\begin{equation}\label{a47}
D^\sigma{}_{\beta_1\cdots\beta_k}:=-k
\,C_{\nu\,(\beta_2\cdots\beta_k}\,\Delta^{\nu\,\sigma}{}_{\beta_1)},
\end{equation}
and the $(\beta_2\cdots\beta_k\,\beta_1)$ means that part which is
symmetric with respect to the indices. Eq. \Ref{a46} can be
rewritten like
\begin{align}\label{a48}
Y^\alpha(t)=\Lambda^\alpha(0)+&\int_0^t\d
t'\;(R^{-1})^\alpha{}_\sigma(t')\,
\,D^\sigma{}_{\beta_1\cdots\beta_k}\nonumber\\
&\times[R^{\beta_1}{}_{\alpha_1}(t')\,
Y^{\alpha_1}(t')]\cdots[R^{\beta_k}{}_{\alpha_k}(t')\,Y^{\alpha_k}(t')],
\end{align}
and from that
\begin{equation}\label{a49}
\Lambda^\alpha(t)=R^\alpha{}_\beta(t)\,\Lambda^\beta(0)+\int_0^t\d
t'\;R^\alpha{}_\sigma(t-t')\,
\,D^\sigma{}_{\beta_1\cdots\beta_k}\,
\Lambda^{\beta_1}(t')\cdots\Lambda^{\beta_k}(t').
\end{equation}
The above expression can be visualized by a set of graphs. Each
graph consists of vertices and directed links. Each vertex has one
outgoing link and $k$ incoming links. Each graph is connected, has
no loops, and has only one outgoing link. $\Lambda^\alpha(t)$ is
the sum of possible such graphs, the values of them are calculated
using the following rules.
\begin{itemize}
\item To each point of the graph (the beginning points, the end
point, and the vertices) is assigned a time. The time
corresponding to the end of a link should not be smaller than the
time corresponding to the beginning of that link. The time
corresponding to the beginnings of the incoming links of the graph
are $0$, and the time corresponding to the end of the outgoing
link of the graph is $t$.
\item To each directed link is assigned a factor $R$, the argument
of which is the time corresponding to the end of the link minus
the time corresponding to the beginning of the link.
\item To each vertex is assigned a factor $D$.
\item To the beginning point of each incoming link of the graph is
assigned a factor $\Lambda(0)$.
\item the value assigned to a graph is the product of the values
assigned to various parts of the graph, integrated over the times
corresponding to the vertices.
\end{itemize}

Using this scheme, one can in principle find $\Lambda(t)$ up to
desired order (number of vertices).
\section{The large time behavior of the system}
The real parts of the eigenvalues of $M$ are nonpositive, and zero
is an eigenvalue of $M$. Assuming that the only eigenvalue with
nonnegative real part is zero, and that this eigenvalue is
nondegenerate, the large time behavior of $R$ is simple. The large
time behavior of $R$ depends on the spectrum of $M$ near zero. If
there is a gap in the spectrum, that is if the supermum of the
real parts of the eigenvalues (apart from zero) is negative, then
\begin{equation}\label{a50}
\lim_{t\to\infty} R^\alpha{}_\beta(t)=u^\alpha\,S_\beta,
\end{equation}
where $u$ is the normalized right eigenvector of $M$ corresponding
to the eigenvalue 0:
\begin{align}\label{a51}
M\,u=&0,\nonumber\\
S_\alpha\,u^\alpha=&1.
\end{align}
If there is no gap in the spectrum of $M$ near the eigenvalue
zero, but still the only eigenvalue of $M$ with zero real part is
zero, and this eigenvalue is nondegenarte, then
\begin{equation}\label{a52}
R^\alpha{}_\beta(t)\sim\left(\frac{t}{\tau}\right)^{-\delta}\,
u^\alpha\,S_\beta,\qquad t\to\infty,
\end{equation}
where $\tau$ and $\delta$ are constants depending on the behavior
of the spectrum of $M$ near zero.

In the case there is a gap in the spectrum, for $t\to\infty$ one
can substitute the right-hand side of \Ref{a50} for $R$ in the
expressions for the graphs, as for \textit{most} of the times, the
argument of $R$ is large. This is equivalent to rewriting
\Ref{a49} as
\begin{equation}\label{a53}
\Lambda^\alpha(t)=u^\alpha\,\left[S_\beta\,\Lambda^\beta(0)+\int_0^t\d
t'\;S_\sigma\, \,D^\sigma{}_{\beta_1\cdots\beta_k}\,
\Lambda^{\beta_1}(t')\cdots\Lambda^{\beta_k}(t')\right],
\end{equation}
or
\begin{equation}\label{a54}
\lambda(t)=S_\beta\,\Lambda^\beta(0)+\int_0^t\d t'\;\mathcal{D}\,
[\lambda(t')]^k,
\end{equation}
where
\begin{equation}\label{a55}
\Lambda^\alpha(t)=:u^\alpha\,\lambda(t),
\end{equation}
and
\begin{equation}\label{a56}
\mathcal{D}:=S_\sigma\,D^\sigma{}_{\beta_1\cdots\beta_k}\,
u^{\beta_1}\cdots u^{\beta_k}.
\end{equation}
It is easy to solve \Ref{a54}. One has
\begin{equation}\label{a57}
\frac{\d\,\lambda}{\d t}=\mathcal{D}\,\lambda^k,
\end{equation}
from which one obtains
\begin{equation}\label{a58}
\lambda(t)=\frac{S_\alpha\,\Lambda^\alpha(0)}{\{1-(k-1)\,\mathcal{D}\,
[S_\beta\,\Lambda^\beta(0)]^{k-1}\,t\}^{1/(k-1)}},
\end{equation}
or
\begin{equation}\label{a59}
\Lambda^\alpha(t)=\frac{u^\alpha\,S_\gamma\,\Lambda^\gamma(0)}{\{1+k\,(k-1)
\,C_{\beta_1\cdots\beta_k}\,u^{\beta_1}\cdots u^{\beta_k}\,
[S_\beta\,\Lambda^\beta(0)]^{k-1}\,t\}^{1/(k-1)}}.
\end{equation}

If there is no gap in the spectrum of $M$ near zero, one has to
substitute \Ref{a52} in \Ref{a49}. It is then seen that
$\Lambda(t)$ is proportional to $u$, for large times. Putting the
ansatz
\begin{equation}\label{a60}
\Lambda^\alpha(t)\sim t^{-\mu}\,u^\alpha,\qquad t\to\infty,
\end{equation}
along with \Ref{a52} in \Ref{a49}, one arrives at
\begin{equation}\label{a61}
t^{-\mu}\sim c\,t^{-\delta}+I(t),
\end{equation}
where
\begin{equation}\label{a62}
I(t)\sim\int_x^{t-y}\d t'\;(t-t')^{-\delta}\,t'^{-k\,\mu}.
\end{equation}
$x$ and $y$ are introduced to ensure that the approximations used
for $R$ and $\Lambda$ are valid in the integration domain. A
dimensional analysis shows that for large $t$,
\begin{equation}\label{a63}
I(t)\sim
c_1\,t^{-\delta}+c_2\,t^{-k\,\mu}+c_3\,t^{1-\delta-k\,\mu},\qquad
t\to\infty.
\end{equation}
So,
\begin{equation}\label{a64}
t^{-\mu}\sim
c'_1\,t^{-\delta}+c'_2\,t^{-k\,\mu}+c'_3\,t^{1-\delta-k\,\mu},\qquad
t\to\infty.
\end{equation}
The meaning of this, is that the two largest exponents entering
this expression should be equal. As $k>1$, the exponent $-k\,\mu$
is smaller than $-\mu$. So it cannot be among the largest
exponents. There remains three possibilities:
\begin{equation}\label{a65}
\mu=\begin{cases} \displaystyle{\frac{1}{k}}, &k\,\delta<1\\ \\
                  \delta, &k\,\delta>1\\ \\
                  \displaystyle{\frac{1-\delta}{k-1}}, &k\,\delta>1\\
\end{cases}.
\end{equation}
It can be shown that the third case does not occur. To see this,
consider the integration corresponding to the a vertex all of its
incoming links are incoming links of a graph. The time dependence
of the integral involved is
\begin{equation}\label{a66}
I'(t)\sim
d_1\,t^{-\delta}+d_2\,t^{-k\,\delta}+d_3\,t^{1-\delta-k\,\delta},
\end{equation}
where $t$ is the time corresponding to the end of the outgoing
link. It is seen that for $k\,\delta>1$, the largest exponent in
the right-hand side is $-\delta$, which shows that the result of
the integration is proportional to $t^{-\delta}$ (for large
times). Repeating this for successive vertices, One finds that all
graphs are proportional to $t^{-\delta}$. So the correct value for
$\mu$ is $\delta$. One can then summarize \Ref{a65} in
\begin{equation}\label{a67}
\mu=\max\left(\delta,\frac{1}{k}\right).
\end{equation}
Defining
\begin{equation}\label{a68}
k_{\mathrm{c}}:=\frac{1}{\delta},
\end{equation}
it is seen that the system exhibits a dynamical phase transition:
for $k>k_{\mathrm{c}}$, annihilation does not enter the relaxation
exponent of the system; while for $k<k_{\mathrm{c}}$, it is the
annihilation which determines the relaxation exponent.

A note is here in order. If instead of \Ref{a60}, one choose an
ansatz that the relaxation of $\Lambda$ is exponential rather than
power law, then the integral on the right-hand side of \Ref{a49}
tends to zero faster than $\Lambda$ itself (as $k>1$), which means
that for large times, only the first term on the right-hand side
of \Ref{a49} determines $\Lambda$, so it should decay
exponentially, which is not the case.

As an example, consider a system consisting of particles of a
single species diffusing on a $d$-dimensional lattice with
symmetric rates. Suppose that there is a $k$-particle annihilation
(double-scaled) as well. The {\em types} of the particles are just
the sites of the lattice, denoted by $\mathbf{x}$ ($d$-tuples of
integers). The matrix $M$ describing the diffusion is then
\begin{equation}\label{a69}
M=\sum_{i=1}^d r_i\,(T_i+T^{-1}_i-2),
\end{equation}
where $T_i$ is the one-site translation in the $i$-th direction.
The eigenvalues of $M$ are
\begin{equation}\label{a70}
E({\boldsymbol{\theta}}):=\sum_{i=1}^d 2\,r_i\,(\cos\theta_i-1).
\end{equation}
For a finite lattice, $\boldsymbol{\theta}$'s are discrete and
there is a gap in spectrum at zero. For an infinite lattice, the
spectrum is continuous at zero, and a steepest-descent study shows
that
\begin{equation}\label{a71}
\delta=\frac{d}{2}.
\end{equation}
This shows that in this case, the system never crosses the
critical point $k=k_{\mathrm{c}}$, as $k_{\mathrm{c}}\leq 2$ and
$k\geq 2$.
\section{Concluding remarks}
A system was investigated consisting of several types of bosonic
particles. By bosonic, it is meant there can be more than one
particle of each type (at each site). In \cite{BH,MPGS,BHPR},
similar systems were investigated and exact results including
phase transitions were obtained for the case of at most two
particle creation or annihilation. In the case investigated here,
there is no creation, but there is $k$-particle annihilation,
where $k$ can be larger than 2. The case of small annihilation
rate, together with large initial number of particles was
investigated in more detail. It was shown that this system
exhibits a dynamical phase transition, which is controlled by $k$
and another parameter related to the rates of one-particle
reactions.


\newpage
\begin{thebibliography}{99}
\bibitem{ScR}  G. M. Sch\"{u}tz; ``Exactly solvable models for many-body
               systems far from equilibrium'' in ``Phase
               transitions and critical phenomena, vol. \textbf{19}'',
               C. Domb \& J. Lebowitz (eds.), (Academic
               Press, London, 2000).
\bibitem{ADHR} F. C. Alcaraz, M. Droz, M. Henkel, \& V. Rittenberg;
               Ann. Phys. \textbf{230} (1994) 250.
\bibitem{KPWH} K. Krebs, M. P. Pfannmuller, B. Wehefritz, \&
               H. Hinrichsen; J. Stat. Phys. \textbf{78}[FS] (1995) 1429.
\bibitem{HS1}  H. Simon; J. Phys. \textbf{A28} (1995) 6585.
\bibitem{PCG}  V. Privman, A. M. R. Cadilhe, \& M. L. Glasser; J. Stat.
               Phys. \textbf{81} (1995) 881.
\bibitem{HOS1} M. Henkel, E. Orlandini, \& G. M. Sch\"utz; J. Phys.
               \textbf{A28} (1995) 6335.
\bibitem{HOS2} M. Henkel, E. Orlandini, \& J. Santos; Ann. of Phys.
               \textbf{259} (1997) 163.
\bibitem{AL}   A. A. Lushnikov; Sov. Phys. JETP \textbf{64} (1986) 811
               [Zh. Eksp. Teor. Fiz. \textbf{91} (1986) 1376].
\bibitem{AKK}  M. Alimohammadi, V. Karimipour, \& M. Khorrami; Phys. Rev.
               \textbf{E57} (1998) 179.
\bibitem{RK2}  F. Roshani \& M. Khorrami; J. Math. Phys. \textbf{43} (2002) 2627.
\bibitem{AKK2} M. Alimohammadi, V. Karimipour, \& M. Khorrami; J. Stat.
               Phys. \textbf{97} (1999) 373.
\bibitem{AM1}  A. Aghamohammadi \& M. Khorrami; Eur. Phys. J. \textbf{B47} (2005)
               583–586 (2005).
\bibitem{KA}   M. Khorrami\& A. Aghamohammadi;  Phys. Rev. \textbf{E70} (2004) 011103.
\bibitem{MG}   D. C. Mattis, \& M. L. Glasser; Rev. Mod. Phys. {\bf 70}, (1998) 979.
\bibitem{THV}  U. C. T\"{a}uber, M. H. Howard, \&  B. P. Vollmayr-Lee; J. Phys.
               \textbf{A38} (2005) R79.
\bibitem{Ta}   U. C. T\"{a}uber; cond-mat/0511743.
\bibitem{BH}   B. Houchmandzadeh; Phys. Rev. \textbf{E66} (2002) 052902.
\bibitem{MPGS} M. Paessens \& G. M. Sch\"{u}tz; J. Phys. \textbf{A37}
               (2004) 4709.
\bibitem{BHPR} F. Baumann, M. Henkel, M. Pleimling, \& J. Richert;
               J. Phys. \textbf{A38} (2005) 6623.
\end{thebibliography}
\end{document}